

\documentclass[conference]{IEEEtran}
\IEEEoverridecommandlockouts
\usepackage{latexsym}
\usepackage{multicol}
\usepackage{algorithmic,cite}
\usepackage{algorithm}
\usepackage{amssymb}
\usepackage{amsmath}
\usepackage{epsfig}
\usepackage{psfig}
\usepackage[dvips]{color}
\usepackage{epsfig}
\usepackage{psfig}
\usepackage{url}
\usepackage{array}
\usepackage{comment}
\usepackage{multirow}
\usepackage{authblk}

\def\BibTeX{{\rm B\kern-.05em{\sc i\kern-.025em b}\kern-.08em
    T\kern-.1667em\lower.7ex\hbox{E}\kern-.125emX}}
\begin{document}
\title{Realistic Indoor Path Loss Modeling for Regular WiFi Operations in India}

\author[1]{Hemant Kumar Rath}
\author[2]{Sumanth Timmadasari}
\author[1]{Bighnaraj Panigrahi}
\author[1]{Anantha Simha}

\affil[1]{TCS Research \& Innovation, India, Email:\{hemant.rath, bighnaraj.panigrahi, anantha.simha\}@tcs.com}
\affil[2]{Dept. of Electrical Engineering, IIT Madras, India, Email:{ee15m074@ee.iitm.ac.in}}


%
\maketitle

\let\oldthefootnote\thefootnote
\renewcommand{\thefootnote}{\fnsymbol{footnote}}
\let\thefootnote\oldthefootnote

\vspace{-2mm}
\begin{abstract}
 Indoor wireless communication using Wireless Fidelity (WiFi) is becoming a major need for the success of Internet of Things (IoT) and cloud robotics in both developed and developing countries. With different operating conditions, interference, obstacles and type of building materials used, it is difficult to predict the path loss components in an indoor environment, which are crucial for the network design. It has been observed that the indoor path loss models proposed for western countries cannot be directly used in Indian scenarios due to variations in building materials utilized, floor plans, etc. In this paper, we have proposed a non-deterministic statistical indoor path loss model- Tata Indoor Path Loss Model (T-IPLM) which can be used for the 2.4 - 2.5 GHz, Industrial Scientific and Medical (ISM) band. To propose and validate, we have conducted several drive tests with different conditions such as busy office premise with obstacles, open office premise, corridor, canteen, and multi-storey office locations, etc. We have also compared T-IPLM with popular path loss models such as ITU-R and Log-distance; T-IPLM matches closely with the drive test results as compared to other models. We believe that T-IPLM model can be used extensively to design accurate indoor communication networks required for regular WiFi communications and deployment and operations of IoT and cloud robotics.

\end{abstract}

\vspace{-1mm}
\section{Introduction}
Wireless communication has matured enough to become the de-facto mode of communication for the last couple of years. With the rise of Internet of Things (IoT) and cloud robotics, more and more smart devices are taking part in the communications. Most of these devices operate in the license free bands or Industrial, Scientific, and Medical (ISM) radio bands. Over the years it has been observed that the  ISM band (2.4 - 2.5 GHz band), popularly known as Wireless Fidelity (WiFi) band is getting congested. This is mainly because millions of devices compete to operate in this license free band resulting in interference. In addition, devices using the above band are mostly used in indoor scenarios which makes their life even more difficult. 

The WiFi standard (IEEE 802.11) specifies the Received Signal Strength Indicator (RSSI) as the measure of the Radio Frequency (RF) energy received by the receiver. Though many researchers have reservations regarding absolute accuracy of RSSI value, it is still being considered as the simplest open loop parameter for received signal strength measurement in practice. Hence, in this paper we have used RSSI value as the measure of signal strength received at a receiver from a transmitter.

It has been observed that based on the transmission power used and antenna gain available, WiFi range is limited to few tens of meters in indoor and few hundreds of meters in outdoor environment. Other than the transmission power and antenna gain, the materials used in the building, the building design pattern, equipments used in the building, floor plan, number of people in the concerned location, etc., also impact heavily on the RSSI value. This is because of the signal loss or path loss occurred in such situations. Accurate path loss needs to be predicted or modeled for optimal deployment of wireless Access Points (APs), deployment of sensors or things, localization of smart devices, smooth control of robots and drones, etc., in both indoor and outdoor environments \cite{PLM24}. Though both of these environments are important, in this paper we concentrate only on the indoor scenario operating in 2.4 - 2.5 GHz band. Note that signal characteristics over 2.X GHz mainly depend upon multi-path propagation along with usual fading and path loss due to distance, interference, shadowing, reflection, refraction and scattering, etc. Further, with the recent introduction of LTE-Unlicensed operation over the ISM bands \cite{laa_aina_2017}, optimal channel assignment in WiFi is complicated. Therefore, we also investigate the impact of interference of smart WiFi devices operating in a laboratory or office environment on other WiFi operations. 

It is to be noted here that there exists various path loss models in literature which are being used today. However we have observed that these path loss models do not match to the drive test data in Indian scenario due to (i) the construction materials used, (ii) scale of smart devices used in the office environment, and (iii) number of users we accommodate in an office environment and their movement patterns, etc. Therefore fresh investigation is desired. In this direction, we attempt to propose a model which can be used to predict the path loss in a typical Indian office environment. This exercise is necessary for optimal deployment of the WiFi Access Points (APs), sensors, robots, etc. Accurate path loss model obtained by this exercise can result in reduced cost of deployment and operation, improved Quality of Services (QoS) in terms of un-interrupted data transmissions, high data rates, optimal transmission power, etc. We also hope that our study can be of use to investigate path loss models and to plan and optimize the next generation networks like 4G/ 5G.

The rest of the paper is organized as follows. In Section \ref{sec_2}, we discuss existing literature and explain path loss models used by the industry today. In Section \ref{sec_3}, we explain data collection method through drive tests and data cleansing techniques used. We discuss our proposed model in Section \ref{sec_4} and evaluate the same in Section \ref{sec_5}. We then discuss future work and conclude this paper in Section \ref{sec_6}.

\section{Literature Survey}\label{sec_2}
Several path loss models applicable to outdoor environment for the band 800 - 1800 MHz are proposed in literature \cite{PLM24, ncc_path_iitg}. Similarly, many indoor propagation models are also proposed in literature. One-slope propagation model \cite{smp2003, capulli2006} is a general path loss model that has been tested in a large number of indoor environments and industrial sites \cite{tanghe2008}. Dual-slope model \cite{rapaport2003} is an extension of one-slope model with better accuracy. Partition model is another such indoor model used for residential and office environments with micro-cell deployment \cite{pahlavan2005}. Time Division Multiple Access (TDMA) based office communication system using pico-cells uses another model as in \cite{akerberg1989}. Another widely used model for indoor signal propagation is the COST-231 multi-wall model \cite{cost1999} which is in the line of outdoor signal propagation model. The average walls model \cite{lloret2004} is proposed to minimize the design efforts of Wireless Local Area Networks (WLANs). These models estimate signal propagation based on complex computations involving vast amount of geometry and terrain data. 


In the literature various statistical models for specific environments have also been proposed. Authors in \cite{plet2010} statistically investigate path loss models for different room categories such as adjacent to transmitter room, non-adjacent, etc., in 14 different houses for the 514 MHz band. In \cite{todd1993}, taking the wall attenuation into account, path loss in four different types of office environments have been determined for the 1.75 GHz and 37.2 GHz bands. Authors in \cite{chrysikos2009, chrysikos2011} have proposed indoor propagation models with lower prediction errors and have analyzed the correctness of their model through drive tests. Their analysis was performed for a site-specific validation of the ITU indoor path loss model such as indoor office environments \cite{chrysikos2009} and indoor airport area. However, their models are very complex and are not generic to be used in all indoor scenarios. In \cite{samir_hameed}, authors have evaluated ITU based indoor path loss model and have examined whether ITU model can be used in office or residential areas. In \cite{lim2004}, Line of Sight (LoS) as well as non-LoS (NLoS) measurements are considered to fit to a one-slope indoor propagation model. The authors have also taken into account the path loss exponents for wall losses in case of NLoS measurements. However, these experiments use high-end circuits and hence are not cost-effective methods for other types of indoor environments.

A multi-wall propagation model has been proposed in \cite{lott2011}. Although this model takes into account the environmental characteristics, it only uses the direct ray between transmitter and receiver. Effect of physical environments on the received signal has been considered in \cite{plets2013} which is used to determine the dominant path between the transmitter and receiver. The authors have modeled the path loss using heuristics by taking into account cumulative wall loss and interaction loss components only.

\section {Path Loss Modeling - Our Approach}\label{sec_3}
From various studies, it is evident that the indoor environment is significantly different from the outdoor environment in many ways. Indoor path loss models need to consider the variations in the floor plans, construction materials used in the building, type and number of office equipments used, number of people working and their movements, scale of smart devices used in the vicinity, etc. Apart from these, multi-path propagation along with usual fading and path loss due to distance, interference, shadowing, reflection, refraction, scattering, and penetration etc., also impact on the received signal characteristics. 


Since we are interested in developing a path loss model for 2.4 - 2.5 GHz band for an indoor scenario, we have conducted several drive tests in a typical office environment with various conditions and constraints. We have selected the operating frequency as per the IEEE 802.11b/g/n standards. It is to be noted here that IEEE 802.11b/g/n has defined 14 overlapping channels as in Table \ref{t:freq_assignments} over the frequency range 2.4 - 2.5 GHz, each with a band of 22 MHz. Out of the 14 overlapping channels, we have randomly selected three channels (Channel 1, 7 and 11 for our experiments) for conducting our drive tests. From the drive test results, we have observed that the popular ITU-R \cite{itu-r} model differs significantly from our drive test data which we explain in the following sections. Therefore, we have attempted to propose a path loss model which can be used for regular operations in an Indian scenario. 

\begin{table}[h!]
\centering
\caption{Frequency and Channel Assignments}
\begin{tabular}{|c|c||c|c|} \hline
\bf{Channel}      &     \bf{Frequency}         &     \bf{Channel}            &     \bf{Frequency}  \\ \hline
1                  & 2.412 GHz          & 8                & 2.447 GHz    \\\hline
2                  & 2.417 GHz          & 9                & 2.452 GHz    \\\hline
3                  & 2.422 GHz          & 10                & 2.457 GHz    \\\hline
4                  & 2.427 GHz          & 11                & 2.462 GHz    \\\hline
5                  & 2.432 GHz          & 12                & 2.467 GHz    \\\hline
6                  & 2.437 GHz          & 13                & 2.472 GHz    \\\hline
7                  & 2.442 GHz          & 14                & 2.484 GHz    \\\hline

\end{tabular}
\label{t:freq_assignments}\vspace{-2mm}
\end{table}

\subsection{Our Approach - Setup}
We have conducted extensive drive tests using the Android App G-Mon \cite{gmon} as in \cite{ncc_path} and a separate Android App designed by our team on Smart phones (Samsung-S3, Coolpad-3 and Google Nexus-4). Both these Apps are used to calibrate and to minimize the error due to application design. In Table \ref{t:system_parameters} we illustrate the operating parameters of our drive tests. 

\begin{table}[h!]
\centering
\caption{Operating Parameters}
\begin{tabular}{|c|c|} \hline
\bf{Operating Parameter} 	       & \bf{Value} \\ \hline
Transmission Power 	  	       & 15 dBm \\ \hline
Frequency Band 	     		       & 2.4 - 2.5 GHz    \\ \hline
Bandwidth			       & 22 MHz \\ \hline
Mobile Height       		       & 2 m \\ \hline
No. of Roofs	      ($h_{roof}$)     & 3-storey buildings \\ \hline
Average area of Office Location	       & 600 sqm \\ \hline

\end{tabular}
\label{t:system_parameters}\vspace{-2mm}
\end{table}

Drive tests are conducted in a multi-storey office with the following options: (i) open office area, (ii) sitting area or cubicles with temporary glass and wooden partitions, (iii) crowded canteen area and (iv) open corridor. We have used the available floor map while conducting our drive tests. We have also conducted the same experiments in other dense residential locations. In addition we have also conducted our experiments in a 3-storey buildings to understand the path loss characteristics in such situations as these kind of situations are very much important from residential point of view. 

\begin{figure}[h!]
\centering
\includegraphics[width=0.48\textwidth]{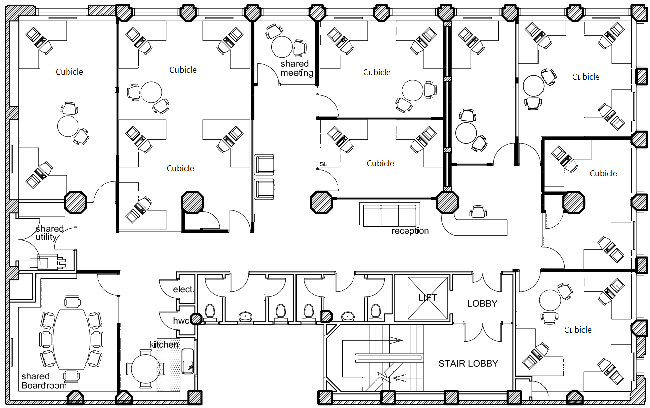}
\caption{Typical Floor Plan}  
\label{fig:drive_test_map}\vspace{-1mm}
\end{figure}

In Fig. \ref{fig:drive_test_map} we illustrate the topology of our drive test scenarios; a snapshot of an office location with multiple cubicles, sitting arrangements, partitions, open space, corridor, etc. For the experiments, we place an WiFi Access Point (AP) at a particular location and then with the floor map in hand, we move to different locations in the office and measure the RSSI values. For each experiment, the location of the transmitter (AP) and receiver (smart phone) are marked on the floor map. This helps us in computing the exact distance between the transmitter and receiver. Moreover, this also helps us in counting the number of walls or obstacles between the Line of Sight (LoS) of the transmitter and receiver, type of walls or obstructions in between, etc., which are crucial for our study. The loss factors for different office walls, pillars and obstacles used in this paper are based on \cite{chrysikos2009, samir_hameed} as in Table \ref{t:obstacle_attenuation}; verified by our experiments as well. 

\begin{table}[h!]
\centering
\caption{Attenuation for Different Obstacles}
\begin{tabular}{|c|c|} \hline
\bf{Obstacle}            &     \bf{Loss (dBm)} \\ \hline
Wooden obstacle                   & 2.67  \\\hline
Concrete wall                   & 2.73 \\ \hline
Pillar (0.6 m $\times$ 0.6 m)               & 6  \\\hline
Glass                         & 4.5  \\\hline
\end{tabular}
\label{t:obstacle_attenuation}\vspace{-1mm}
\end{table}

\subsection{Data Collection and Cleansing}
\begin{figure}[h!]
\centering
\includegraphics[width=0.49\textwidth]{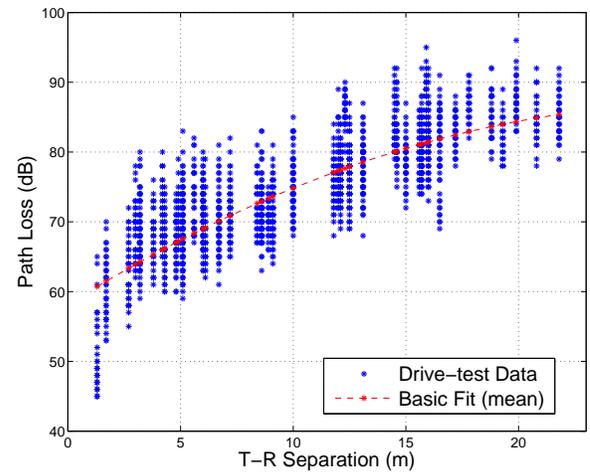}
\caption{Snapshot of Path Loss values in an Indoor environment}  
\label{fig:drive_test_result}\vspace{-1mm}
\end{figure}

\begin{figure}[h!]
\centering
\includegraphics[width=0.49\textwidth]{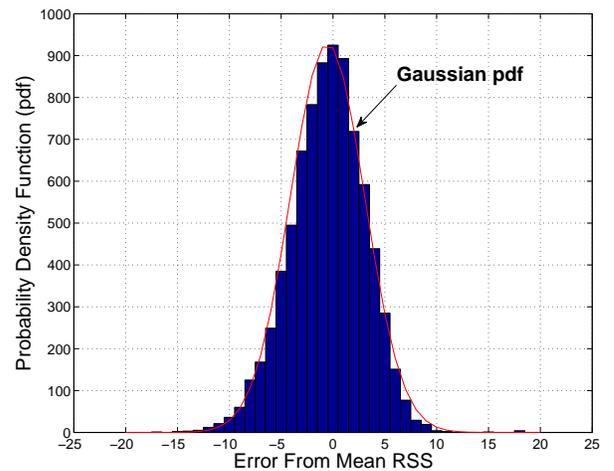}
\caption{Statistical Analysis of Path Loss Values}  
\label{path_loss_stat1}\vspace{-1mm}
\end{figure}

Using the Android-Apps as mentioned earlier in this paper, we have repeated the same experiment for about a week and have collected the RSSI values at multiple locations (with different AP and smart phone placement) in different times of the day (early office hours, lunch hours, early evening and late evening, etc.) and in different crowd scenarios (empty, half and fully occupied). Fig. \ref{fig:drive_test_result} illustrates the path loss values at different transmitter-receiver distances. We have noted the min, max and the mean values of the path loss values being observed at each location. From this figure, we have observed that mean or average path loss value measured can be used as an indicator for path loss modelling. We have also plotted the histogram of the error between the mean, and the logged path loss values in Fig.  \ref{path_loss_stat1}. From this figure, we have observed that the histogram of the error matches to Gaussian distribution with 0.5 dB as mean and 3.58 as the standard deviation. Since the standard deviation and mean of the error are low, mean of the observed drive test results can be used for path loss modeling (Fig. \ref{fig:drive_test_result}). 

\section{Proposed Model: T-IPLM}\label{sec_4}
Our proposed model is an empirical non-deterministic statistical model named as Tata Indoor Path Loss Model (T-IPLM). As in \cite{ncc_path_iitg}, we also divide the path loss values into two categories: (i) fixed path loss value as a function of frequency of transmission and (ii) empirical loss due to variations in the signal path impacted due to obstacles, diffraction, multi-path, etc. The first segment is being contributed by the LoS and NLoS nature of the transmission and is an direct extension of ITU-R model. For the sake of completeness, ITU-R model is expressed as:

\begin{equation}\label{eq:pathloss-itur}
  \begin{split}
	  PL_{ITU-R} (dB) = & 20 \times \log10(f) + N \times \log10(d) \\ & + P_f(n) -28,
	  \vspace{-2mm}\end{split}
\end{equation} 

\noindent where $PL_{ITU-R}$ is the path loss value in dB, $f$ is the operating frequency in MHz, $d$ is the distance between the transmitter and receiver in meters, $N$ is the distance power loss coefficient, $P_f(n)$ is the floor loss prediction factor and $n$ is the number of floors between transmitter and receiver. Empirical value of $N$ is used as 30, 28 and 22 for office, residential and commercial areas respectively. We have also used Log-distance model for comparison:

\begin{equation}\label{eq:pathloss-log}
  \begin{split}
	  PL_{Log} (dB) = & 20 \times \log10 \bigg(\frac{4 \times \pi \times d_0}{\lambda}\bigg)  \\ & + 10\times {\gamma} \times \log\frac{d}{d_0},
	  \vspace{-2mm}\end{split}
\end{equation} 

\noindent where $\lambda$ is the operating wavelength, $d_0$ is the reference distance (1 m) and $\gamma$ is the path loss exponent. Using curve fitting mechanisms and taking a clue from the ITU-R model, we propose our path loss model - Tata Indoor Path Loss Model (T-IPLM) as:

\begin{equation}\label{eq:pathloss-itur}
  \begin{split}
	  PL_{T-IPLM} (dB) = & 20 \times \log10(f) + N_T \times \log10(d) \\ & + \sum_w L_w + FAF -20,
	  \vspace{-2mm}\end{split}
\end{equation} 

\noindent where $N_T$ is the power loss coefficient due to distance, $L_w$ is the LoS loss factor of walls (glass or wooden or temporary partitions) between the transmitter and receiver and $\sum_w L_w$ is the total loss only due to walls and $FAF$ is the floor attenuation factor. 

Path loss component related to frequency is as per the ITU-R model. We have divided other path loss components into multiple segments: (i) loss due to distance only, (ii) loss due to obstructions such as walls, partitions, and (iii) loss due to ceilings in multi-storey buildings. We have used curve fitting techniques to obtain the constant parameter 20 used in our model. $N_T$ values are obtained from several drive tests conducted in open office area. $L_w$ values are obtained by conducting drive tests in busy office area with multiple obstructions between the transmitter and the receiver. We have also observed that the value of $N_T$ differs for different channel of operations. Based on several rounds of drive tests, we have obtained the value of $N_T$ for Channel 1, 7 and 11 as in Table \ref{t:N_values}.

\begin{table}[h!]
\centering
\caption{$N_T$ Value for Different Number of Obstacles}
\begin{tabular}{|c|c||c|c||c|c|} \hline
\multicolumn{1}{|c}\bf{Channel 1} & \multicolumn{2}{c} \bf{Channel 7} & \multicolumn{3}{c} \bf{Channel 11}\\\hline
\bf{\# Obstacles}  & \bf{$N_T$}     & \bf{\# Obstacles}     & \bf{$N_T$}      & \bf{\# Obstacles}     & \bf{$N_T$}\\ \hline
1            & 31.1          & 1        & 32.9            & 1        & 29.3    \\\hline
2            & 30.1         & 2        & 28.5            & 2        & 28.4     \\\hline
3            & 31.8          & 3        & 26.7            & 3        & 27    \\\hline
4            & 31.2          & 4        & 29.1            & 4        & 28    \\\hline
5            & 31.3          & 5        & 27.4            & 5        & 28.4    \\\hline

\end{tabular}
\label{t:N_values}\vspace{-1mm}
\end{table}

Value of $L_w$ for different types of walls is as per Table \ref{t:obstacle_attenuation}. For multi-storey buildings, using several experiments we have obtained FAF values as mentioned in Table \ref{t:floor_attenuation}; $FAF=0$, if the transmitter and receiver are in the same floor and  $FAF > 0$, otherwise. These values are specific to concrete ceilings with usual extra false ceilings (Poly Vinyl Chloride: PVC-based) and tile or marble flooring, which are typically used in India. 

\begin{table}[h!]
\centering
\caption{Floor wise Attenuation Factors}
\begin{tabular}{|c|c|} \hline
\bf{Scenario}                     & \bf{FAF (dBm)} \\ \hline
1 Floor above                     & 21  \\\hline
2 Floors above                    & 33 \\ \hline
3 Floors above                    & 40  \\\hline
1 Floor below                     & 21  \\\hline
2 Floors below                    & 36 \\ \hline

\end{tabular}
\label{t:floor_attenuation}\vspace{-2mm}
\end{table}

To understand the correctness of our proposed model we now explain the experimental results for various scenarios. We have also compared the performance of our model with the popular ITU-R and Log-distance models.  

\section{Performance Evaluation}\label{sec_5}
\begin{figure}[h!]
\centering
\includegraphics[width=0.49\textwidth]{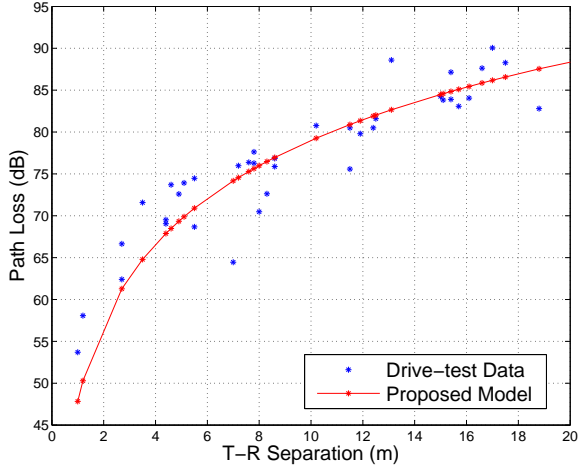}
\caption{Path Loss in Busy Office Premises: Channel 1}  
\label{path_loss_closed1}\vspace{-1mm}
\end{figure}

\begin{figure}[h!]
\centering
\includegraphics[width=0.49\textwidth]{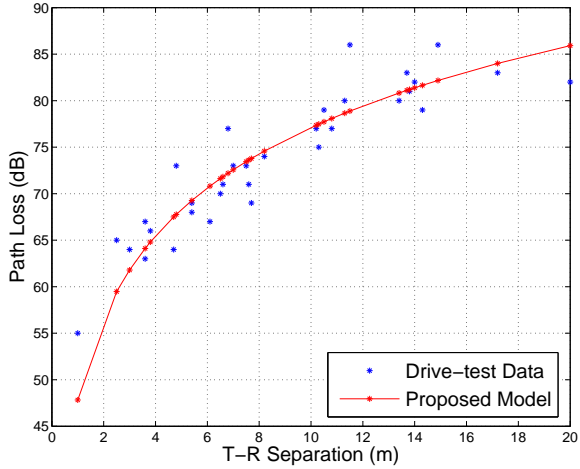}
\caption{Path Loss in Busy Office Premises: Channel 7}  
\label{path_loss_closed7}\vspace{-1mm}
\end{figure}

To evaluate the nature and correctness of our model T-IPLM, we have conducted drive tests for one more week (i.e., one week drive test for modeling and one week for validation and comparison) and compared the proposed model with the experimental data. Fig. \ref{path_loss_closed1} and Fig. \ref{path_loss_closed7} illustrate the correctness of our model with the experimental data; busy office premises for different operating channels (Channel 1 and Channel 7). It is to be noted that while conducting the drive tests, we have not only used other WiFi APs as the interferes operating in the same channel, but also used additional WiFi dongles to create interference. From these figures, we observe that path loss values obtained by our proposed model is close to the average path loss values obtained from the drive test data. 

\begin{figure}[h!]
\centering
\includegraphics[width=0.49\textwidth]{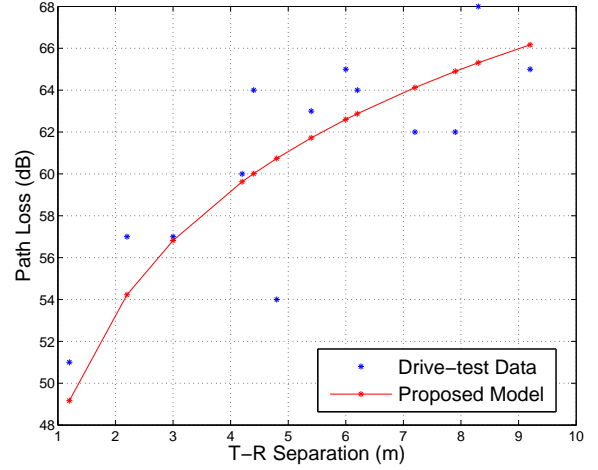}
\caption{Path Loss in Open Space: Channel 1}  
\label{path_loss_open1}\vspace{-1mm}
\end{figure}

\begin{figure}[h!]
\centering
\includegraphics[width=0.49\textwidth]{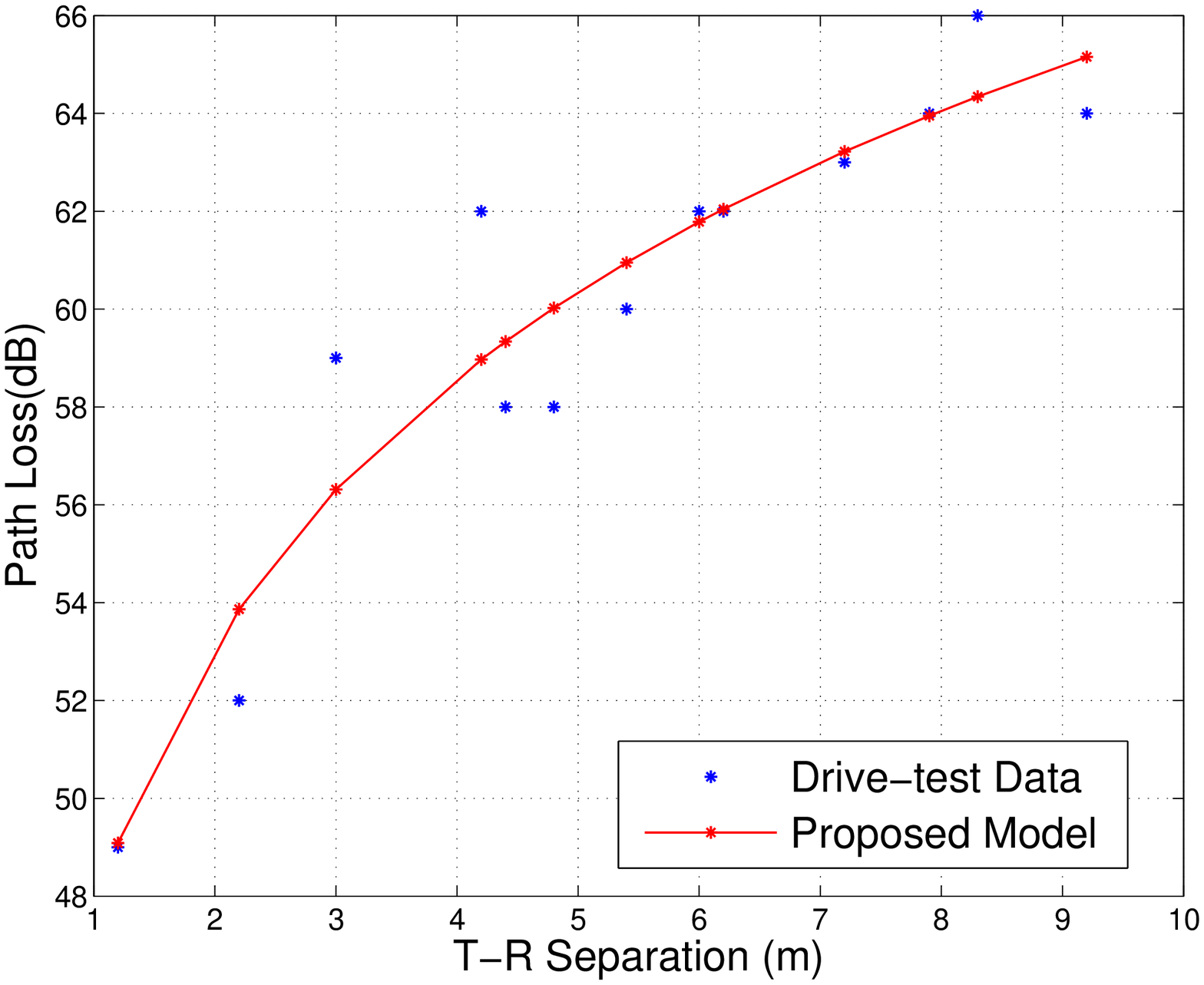}
\caption{Path Loss in Open Space: Channel 7}  
\label{path_loss_open7}\vspace{-1mm}
\end{figure}

\begin{figure}[h!]
\centering
\includegraphics[width=0.49\textwidth]{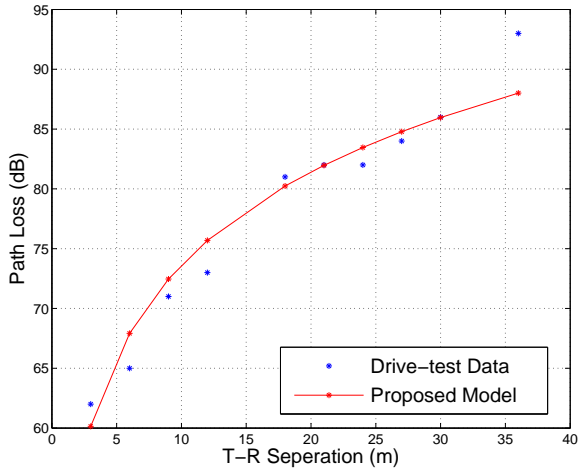}
\caption{Path Loss in Open Corridor}  
\label{path_loss_corridor}\vspace{-1mm}
\end{figure}

Similar to busy office premises, in Fig. \ref{path_loss_open1} and Fig. \ref{path_loss_open7}, we validate our model for a relatively open space\footnote{Open Space: Area where there is no major obstacle for about 10-15 m.} scenario in an office premise (Channel 1 and Channel 7). For the open space, we observe that the value of $N_T$ differs significantly from the values obtained in Table \ref{t:N_values}; for Channel 1: $N_T=19.2$, for Channel 7: $N_T=18$ and Channel 11: $N_T=17.3$. Similar to our earlier observations for busy office premises, our model is closely matching to the observed path loss values in open space as well. Apart from the busy and open office premises, we have also considered open corridors to validate our model. As illustrated in Fig. \ref{path_loss_corridor} our model also matches to the drive test data. Value of $N_T$ obtained from our drive test data for open corridors is around 25.8. From both the open space office premises experiment and open corridors experiment, we have observed that path loss component $N_T$ is significantly more for corridors as compared to open office premises. This is mainly because of the narrowness of the corridors and multi-path propagation and reflections that become inevitable in corridors.

\begin{figure}[h!]
\centering
\includegraphics[width=0.49\textwidth]{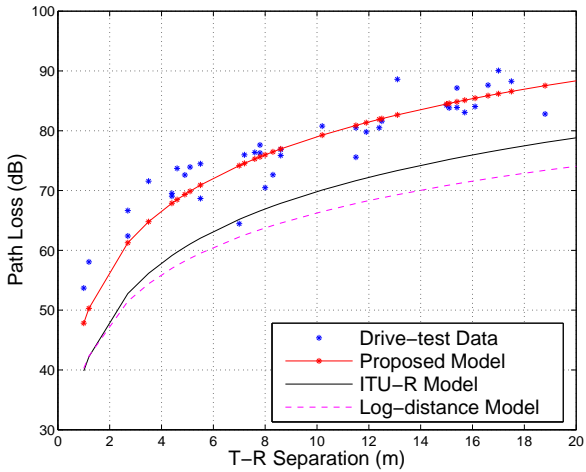}
\caption{Comparison of Our Proposed Model with Others}  
\label{path_loss_comparison1}\vspace{-1mm}
\end{figure}

We have also compared our proposed model T-IPLM with that of ITU-R and Log-distance model in Fig. \ref{path_loss_comparison1} (Channel 1, busy office premise). From this figure, we have observed that our model matches closer with the average drive test data as compared to other models. Due to page restrictions other comparison plots are not presented in this paper.

\subsection{Discussions}
From drive test data, proposed model and the existing models (ITU-R and Log-distance) we have observed that:

\begin{itemize}

\item The Mean Square Error (MSE) between the mean observed path loss and our proposed path loss values varies between 1.49 - 3.6; minimum in the corridor and maximum in the busy office premises. 

\item The MSE value obtained for T-IPLM (Channel 1, busy office premise) is around 3.6, whereas it is 10.3 for ITU-R and 13.2 for Log-distance model. Moreover, there is a significant lag between the mean observed values and the computed ITU-R and Log-distance values as compared to T-IPLM.

\item From the above observations, we believe that T-IPLM is statistically a better model as compared to ITU-R and Log-distance models. We therefore argue that T-IPLM can be used as a better estimator of path loss for indoor Indian environment for the band 2.4 - 2.5 GHz.

\end{itemize}

\section{Conclusions and Future Work}\label{sec_6}
In this paper we have proposed an indoor path loss model called Tata Indoor Path Loss Model (T-IPLM) which can be used for regular WiFi operations for the band 2.4 - 2.5 GHz in India. Based on several drive tests conducted in a typical office environment, we have formulated a mathematical model which can be used in - indoor open area, open corridors, crowded office premises, multi-storey buildings, etc. We have also compared T-IPLM with popular path loss models used in practice such as ITU-R and Log-distance model and have demonstrated the correctness of our model. This model can be suitably extended to other countries through rigorous experiments. Due to its adaptive nature, T-IPLM can be used for regular indoor IoT deployment and robotics operating in 2.4 - 2.5 GHz. As a part of future work, we intend to work on path loss models for other frequency bands such as 5.8 GHz for regular WiFi and LTE-Unlicensed operation, 2.4 - 2.5 GHz for regular WiFi communication in mines, tunnels etc.
\bibliographystyle{ieeetr}

\bibliography{ulscheduler}

\end{document}